# Exploring the Nuances of Designing (with/for) Artificial Intelligence
Niya Stoimenova, Rebecca Price

For all of the technology advancements since the creation of the internet, the infant stages of a true digital economy are only just being realized. A fundamental shift in how society operates is approaching, driven by advances in artificial intelligence (AI) (often referred to interchangeably as one of its branches—machine learning, ML). There are complexities to incorporating AI into products, services, and systems requiring attention of the design discipline. Although progress of AI is compelling, there is an observed comparative lack of discourse across the design discipline regarding this topic.

    The aim of this article is to shed light on the deeper consequences of AI development for design. Our rationale is to avoid repeating hype-associated dialogue of utopian/dystopian technology futures. We undertake scene-setting and review major perspectives within the AI landscape to frame this article. Our emphasis then moves to presenting and discussing technology-related developments in relation to the design movement. We identify that the problems addressed by AI-powered solutions are becoming increasingly broad in scope and inherently wicked. Practical examples illustrate this point. Yet AI solutions remain riddled with biases that when poorly conceived, can cause more damage than good.

    A methodological gap emerges from our analysis. We identify that there are fragments of methodological readiness for AI in the discipline, but these are not yet connected to meet the nuances of designing with this new subject matter. Nuances such as the challenge of designing to ensure AI-powered artifacts remain safe even as the utility they deliver evolves over time—both through instructional and machine learning (ML) prompted by the user's interactions and input from a broader system. This article brings together the methodological developments of participatory and human-computer interaction design to consider a new approach for designing with AI.

## Setting the Scene

In 2016, an experimental driverless vehicle developed by researchers at Nvidia performed considerably better than its competitors (Alphabet, Uber, Tesla). Unlike other autonomous vehicles, it learned how to behave on the road by watching a human instead of being provided instructions on how to drive. However, because of the standard way such algorithms are devised, the vehicle's reasoning processes were largely opaque—a mystery even its developers struggled to untangle.[1] To address the problem, Mariusz Bojarski and colleagues at Nvidia, New York University, and Google Research developed a simple method for highlighting the parts of the image the algorithm pays attention to.[2] However, the rationale behind why these parts of the image were highlighted remained largely unknown.

## The AI Landscape

The development of AI is expected to occur in three stages: narrow (weak) AI, general (strong) AI, and artificial superintelligence or intelligence amplification.[3] Narrow AI, currently achieved and believed by many scientists to be the only possible incarnation of intelligent machines, is bound to a specific field and is incapable of performing tasks outside a preprogramed scope. Some examples are the way Netflix and Spotify generate recommendations, the use of chatbots to address customer inquiries, and the way Facebook decides what to curate in the user's newsfeed (although contentiously). Although some widely publicized AI implementations tackle more general tasks, such as driving a car (Tesla's autopilot) or generating music (IBM's Watson), these examples are still considered a coordination of several narrow AIs.

The second speculative stage is general AI, which Shulman and Bostrom, researchers at the Future of Humanity Institute at University of Oxford define as "systems which match or exceed the [intelligence] of humans in virtually all domains of interest."[4] A growing number of renowned scientists, philosophers, and forecasters predict the creation of general AI by mid-twenty-first century. It is their belief that accelerating progress in hardware, artificial intelligence, robotics, genetic engineering, and nanotechnology makes this timeframe achievable.[5] The advent of general AI could eventually trigger an event called technological singularity.[6] Currently, many scholars and practitioners have argued that the growth of machine intelligence is likely to radically affect civilization.[7]

Due to its controversial nature, the notion of technological singularity has prompted two possible dichotomous scenarios for the future. The first predicts the emergence of artificial super intelligent agents with in-telligence reaching far beyond the collective capabilities of all renowned human experts in knowledge fields, including scientific creativity, general wisdom, and social skills, as suggested by Bostrom and Yudkowsky.[8] This scenario could also be seen as somewhat akin to Georg Hegel's description of the ascent of human culture to an ideal point of absolute knowing.[9] Many prominent entrepreneurs, scientists, and philosophers, such as Bill Gates, Stephen Hawking, Elon Musk, and Sam Harris believe this scenario could lead to human extinction.

---

1 Will Knight, "The Dark Secret at the Heart of AI," *MIT Technology Review* (April 11, 2017).
2 Bojarski Mariusz, Philip Yeres, Anna Choromanaska, Krzysztof Choromanski, Bernhard Firner, Lawrence Jackel, and Urse Muller, "Explaining How a Deep Neural Network Trained with End-to-End Learning Steers a Car," CoRR (2017).
3 Christopher Noessel, *Designing Agentive Technology: AI That Works for People* (New York: Rosenfeld Media, 2017).
4 Carl Shulman and Nick Bostrom, "How Hard Is Artificial Intelligence? Evolutionary Arguments and Selection Effects," *Journal of Consciousness Studies* 19, nos. 7–8 (2012): 103–30.
5 Vincent C. Müller and Nick Bostrom, "Future Progress in Artificial Intelligence: A Survey of Expert Opinion," in *Fundamental Issues of Artificial Intelligence* (Cham, Switzerland: Springer, 2016), 555–72.
6 Amnon H. Eden, James H. Moor, Jonny Soraker, and Eric Steinhart, *Singularity Hypotheses: A Scientific and Philosophical Assessment* (Berlin: Springer, 2012); Vernor Vinge, "The Coming Technological Singularity: How to Survive in the Post-human Era," *Whole Earth Review* (1993).
7 Gregory S. Paul and Earl Cox, *Beyond Humanity: Cyberevolution and Future Minds* (Rockland, MA: Charles River Media, 1996); Damien Broderick, *The Spike: How Our Lives Are Being Transformed by Rapidly Advancing Technologies* (New York: Tom Doherty Associates, 2002); Ray Kurzweil, *The Singularity Is Near* (London: Viking, 2010); Iyad Rahwan, Manuel Cebrian, Nick Obradovich, Josh Bongard, Jean-François Bonnefon, Cynthia Breazeal, Jacob W. Crandall et al. "Machine Behaviour," *Nature* 568, no. 7753 (2019): 477–86.

The second scenario describes the emergence of a post-human race, evolved with amplification of human cognitive capabilities.[10] The new race would theoretically overcome current physical and mental limitations and, in the most extreme sense, conquer disease, aging, and even death.[11] For many, this type of forecasting is uncomfortable and even a cause for cognitive dissonance toward the topic. But multiple companies are already working to explore this scenario. Neuralink is developing a neural lace that is said to improve bandwidth of communication between human brains and machines, Facebook announced interest in working on enabling people to type with their thoughts,[12] and start-up CTRL-Labs demonstrated a prototype that allows users to interact with a machine by sending signals through their spinal column.[13] These technology futures have been confined to the realms of science fiction to date—but now confront us contentiously.

It is plausible that we may never enter the latter two stages of AI development. What is already occurring is the integration of narrow AIs into everyday life with new products, services, and systems.[14] From music and news recommended by algorithms to the way devices are unlocked with facial recognition, many services and devices regularly mine user behavior and contextual data to tailor highly personalized services and experiences. An object like a refrigerator now understands the user's diet by sensing its own shelf contents and may even order weekly groceries. Collectively, a network of home appliances supports the occupants in living a fulfilling lifestyle. Identity becomes deeply enmeshed in a nonbiological matrix of machines, tools, codes, and semi-intelligent daily objects.[15] Such technologies are becoming less like tools and more like part of an extended mental apparatus of the person.[16] The maturity and prevalence of this technology has catalyzed the notion that ML is the new user experience (UX). That is, ML will be the most important way to improve user experience.[17] Such depth of knowledge initiates the transition from highly personalized experiences to personalized realities highlighting the interplay between expert and everyday ideas of appropriate practice. This is not just a practical issue concerning what works and what doesn't under specific circumstances. It holds profoundly ethical implications—who determines what artificial doings that deeply affect society can do?[18] To unpack this conundrum, we need to understand the dual nature of AI (ML) problems, even in a narrow state.

**Duality of Problems in Narrow AI**
In April 2018 Ryen White and coauthors from Microsoft Research and Duke University published a paper reporting their initial attempts to create a "simple scalable test that can be used for screening of Parkinson's disease in the community or at home."[19] The researchers used longitudinal log data from Microsoft's search engine, Bing, to look into the presence and frequency of symptom-related query terms; motor symptoms, such as the speed, direction, and tremors of cursor movements; and presence of risk factors. Despite still being in a testing phase, their model showed promise in detecting a disease that has a current clinical early diagnosis accuracy of approximately 80 percent.[20] Similarly, Stanford University researchers devised an algorithm that performed better than radiologists in detecting pneumonia from front-view chest X-ray images.[21] Scientists from Google, Harvard University, and University of Connecticut created an algorithm that can forecast the aftershock locations of earthquakes and identify "physical quantities that may control earthquake triggering during the most active part of the seismic cycle."[22] These examples are complemented by advances in self-driving cars and the accompanying new concepts of mobility; anti-aging efforts (e.g., Alphabet's Calico), and optimizing agriculture (e.g., AgriSight's FarmLogs). Although these models are promising, the consensus in the scientific community is that these initial use cases require further investigation over time.

Such ML applications are used to automatically detect patterns in data and use these to predict future data.[23] As such, they inherently solve what Rittel and Webber call tame problems.[24] Tame problems involve an enumerable set of solutions, clear rules, and binary decision mechanisms (true or false). However, what is devised to solve the tame problem of predicting data patterns transcends its initial boundaries and begins to affect the larger social system in which it is situated. At this point, the tame solution enters and interacts in a wicked environment. According to Rittel and Webber, wicked problems are a "class of social system problems which are ill-formulated" and defined by confusing information, multiple clients and decision makers with conflicting values, and significant ramifications.

To exemplify this transition of AI from tame to wicked problem, we return to the work of White and colleagues. Their focus is accurately predicting the presence of factors that potentially signal

---

18  Lenneke Kuijer and Elisa Giaccardi, "Co-performance: Conceptualizing the Role of Artificial Agency in the Design of Everyday Life," *Proceedings of the 2018 CHI Conference on Human Factors in Computing Systems* (ACM, 2018), 125.
19  Ryen W. White, P. Murali Doraiswamy, and Eric Horvitz, "Detecting Neurodegenerative Disorders from Web Search Signals," *Digital Medicine* 1, no. 1 (2018): 8.
20  Ibid.
21  Pranav Rajpurkar, Jeremy Irvin, Kaylie Zhu, Brandon Yang, Hershel Mehta, Tony Duan, Daisy Ding, et al., "Chexnet: Radiologist-Level Pneumonia Detection on Chest C-Rays with Deep Learning," Working paper, arXiv:1711.05225v3, Cornell University Library (2017).
22  Phoebe M. R. DeVries, Fernanda Viégas, Martin Wattenberg, and Brendan J. Meade, "Deep Learning of Aftershock Patterns Following Large Earthquakes," *Nature* 560, no. 7720 (2018): 632.
23  Kevin Murphy, *Machine Learning: A Probabilistic Perspective* (Cambridge, MA: MIT Press, 2012).
24  Horst W. J. Rittel and Melvin M. Webber, "Wicked Problems," *Man-made Futures* 26, no. 1 (1974): 272–80.

Parkinson's disease. However, their ambition to create a test for community and home use inevitably posits a social problem. The project encounters sensitivities concerning quality of life and mortality. The project responds to the problem formulation of "identify and diagnose." Yet the information delivery of this type of solution requires careful consideration of perspectives in allied health systems. How does the solution deliver a diagnosis? What role do doctors play when diagnosis is outsourced? How does the solution connect the diagnosed within the allied health service? How might the family require support after a diagnosis? Consequently, there is no enumerable set of potential solutions nor a well-described set of permissible operations. Delivering a false diagnosis or delivering a diagnosis insensitively could cause significant distress to the community.

In this arrangement, a solution is under pressure to surpass true or false criteria (identify and diagnose) and move to an ethical evaluation—right and wrong. Yet as Rittel and Webber note, because "many parties are equally equipped and interested to judge the solution, none has [sic] the power to set formal decisions rules to determine correctness."[25] Finally, there is no immediate and ultimate test of a solution to the problem that ensures positive impact. Every solution to this problem is a "one-shot operation" and the epitome of a wicked problem described by Rittel and Webber.

Two prominent examples highlight how a tame problem–solving algorithm may cause distress when interacting with the broader social system. First, consider the algorithms Google and Facebook use to rank pages and show content. In 2016 these algorithms were used as a tool for mass misinformation and manipulation, resulting in the infamous phrase "fake news." Despite both companies' efforts to prevent such situations after the US presidential election in 2016, the spread of fake news after the mass shooting in Las Vegas in October 2017 proved very difficult to prevent.[26] Second, consider the case of ProPublica. In 2016 ProPublica ran an experiment with an algorithm widely used in the US judicial system. The algorithm exhibited racial biases and proved to be highly inaccurate, resulting in many falsely identified defendants, described in the work of Julia Angwin and colleagues.[27] In both cases, the ML-powered solutions were driven by data availability and learner performance rather than deliberate vision.[28] Thus, they failed to account for users in different scenarios despite being widely adopted in many newly introduced consumer goods and services.[29] James Guszcza and colleagues from Deloitte in partnership with MIT note, "It is by now abundantly clear that, left unchecked, AI algorithms embedded in digital and social technologies can encode societal biases, accelerate the spread of rumors and disinformation, amplify echo chambers of public opinion, hijack our attention, and even impair our mental wellbeing."[30] Whereas Guszcza and colleagues emphasize reactive auditing of implemented algorithms to correct wrongful activity, we see great value in methodological developments in design that stem from a tradition of dealing with wicked problems. Design can lead to a more robust cycle of developing of AI-powered solutions.

**Views on Design with AI/ML**

Discussion in the design discipline regarding the implications of AI on design is limited, with scholars from human–computer interaction (HCI) taking leadership so far. Discourse concerns the best paradigm for approaching the new era with two dichotomous views apparent: (1) human-centeredness and (2) co-performance. These viewpoints are described further.

  A large part of the developments in the history of the HCI field are geared toward designing for human-centeredness.[31] Human-centered design advocates modeling users' natural behavior in interface design so that it becomes intuitive and easier to learn and has fewer performance errors.[32] Fundamentally, it is "an affirmation of human dignity … and an ongoing search for what can be done to support and strengthen the dignity of human beings as they act out their lives in varied social, economic, political, and cultural circumstances."[33] This view is widely adopted in academia and practice. For instance, in July 2017, Google officially established their People in AI Research (PAIR) initiative, which aims to conduct "fundamental research, invent new technology, and create frameworks for design in order to drive a humanistic approach to artificial intelligence."[34] The initiative regularly releases projects and resources that help designers become acquainted with the possibilities AI offers. Another prominent design company that announced their intention is IDEO. In late 2017, IDEO acquired the data science company Datascope with the ambition to "create an offering we're calling D4AI: Design for Augmented Intelligence, which will be able to extend the capabilities of humans in a way that feels natural to them."[35] Microsoft is also already applying their inclusive design principles to AI development.[36]

  A recently introduced paradigm on the role design can play in the creation of solutions powered by ML is that of co-performance. Unlike the human-centeredness perspective, the artifact in what Kuijer and Giaccardi[37] call co-performance is seen as "capable of learning and performing a social practice together with people." There is a direct link between decisions made in the design process and use practices carried out after. The locus of design is thus shifted toward solutions that allow for a recursive relation between design and use, allowing more room for evolving complementary capabilities and doings. Kuijer and Giaccardi argue that the concept of co-performance shows potential to be developed into a range of design approaches and tools that can help designers of computational artifacts acknowledge that appropriate practice varies over situations and changes over time. The project Resourceful Ageing, funded by the Netherlands Organisation for Scientific Research, demonstrates how ethnographic research paired with insights from co-performing household things offers a way to design with ML. The project led to the design of data-enabled products, service propositions, and simple interventions that promoted the vision that aging is an achievement and should be celebrated.[38] The question remains: how might these products, services, and systems grow with users in surprising and delightful ways over time without diverging towards harm?

**Infrastructure as a Necessary Design Material**

We have introduced the notion of ML-powered solutions inadvertently transitioning into the area of wicked problems and design as a discipline addressing this transition. However, choosing to view the problems ML addresses deliberately or inadvertently as tame or wicked means that an important aspect of the dual nature of AI-powered solutions is overlooked. To fully mitigate undesirable outcomes in the narrow state of AI or its future incarnations, we must take a holistic view.

We turn to the construct of infrastructure as a means of simultaneously addressing tame and societal issues in the context of AI. Prominently discussed by Star and Ruhleder,[39] infrastructure is a combination of interrelated social, technical, and organizational arrangements.[40] As such, it is a complex matrix of objects and standards without absolute boundaries or a priori definition.[41] The construct is familiar to the design discipline. An active area in the field of participatory design has been built around the notion of infrastructuring.[42] Erling Björgvinsson and colleagues define it as an opportunity to extend design toward an open-ended, long-term, and continuous process.[43] The approach facilitates the emergence of new design opportunities by deliberately designing indeterminacy and incompleteness into solutions.[44] This leaves space for unanticipated events and performances so that solutions designed at project time can design boundary objects (infrastructure) supportive of future design (at use time), essentially creating a chain of one design after another.

This notion of design after design is well suited for the nature of AI-powered solutions where users regularly develop their own functionality,[45] such as teaching an algorithm their preferences to create highly personalized experiences. By choosing which songs to play, for how long, and how often, the user implicitly instructs a platform like Spotify to adapt to their needs—designing after it has been designed. This symbiosis between tame and wicked problems complemented by inadvertent designing after design can be made explicit by purposefully building infrastructure. To realize this proposition, we deconstruct the three main dimensions of infrastructure identified in the literature and offer ways to build structures that can support the notion of design after design. These elements are social, organizational, and technological.

---

38 Iohanna Nicenboim, Elisa Giaccardi, and Lenneke Kuijer, "Designing Connected Resources for Older People," *Proceedings of Designing Interactive Systems Conference 2018* (ACM Press, 2018).

39 Susan Leigh Star and Karen Ruhleder, "Steps Toward an Ecology of Infrastructure: Design and Access for Large Information Spaces," *Information Systems Research* 7, no. 1 (1996): 111–34.

40 Susanne Bødker, Christian Dindler, and Ole Sejer Iversen, "Tying Knots: Participatory Infrastructuring at Work," *Computer Supported Cooperative Work* 26, nos. 1–2 (2017): 245–73.

41 Susan Leigh Star, "Infrastructure and Ethnographic Practice: Working on the Fringes," *Scandinavian Journal of Information Systems* 14, no. 2 (2002): 6.

42 Helena Karasti, "Infrastructuring in Participatory Design," *Proceedings of the 13th Participatory Design Conference: Research Papers* (ACM, 2014), 1:141–50.

43 Erling Björgvinsson, Pelle Ehn, and Pers-Anders Hillgren, "Design Things and Design Thinking: Contemporary Participatory Design Challenges," *Design Issues* 28, no. 3 (Summer 2012): 101–16.

44 Pers-Anders Hillgren, Anna Seravalli, and Anders Emilson, "Prototyping and Infrastructuring in Design for Social Innovation," *CoDesign* 7, nos. 3–4 (2011): 169–83.

45 Björgvinsson, Ehn, and Hillgren, "Design Things and Design Thinking."

**Social**

The design discipline is already well equipped to deal with this element, addressing it from either human-centeredness or co-performance perspective. However, progress is still needed to ensure AI-powered solutions remain aligned with human values. The efforts of Microsoft, Google, IDEO, and many other design agencies and scholars are geared toward this. A series of use cases by Microsoft incorporating inclusive design explored the requirements of an AI chatbot in various scenarios where children interact with the technology,[46] and they discerned how design could be applied to identify and reduce bias—a practicality advocated by Jeanne Liedtka.[47] Joyce Chou and colleagues at Microsoft argue that design as an activity plays a critical role in developing an ethical framework based on requirements for all potential users, not just a lead user or set of users.[48] Moreover, design encourages a deeper consideration of a user's desires and emotions in context—with intention to build knowledge beyond preliminary requirements. Where physical prototypes are not possible, Lloyd notes the power of imagination as a means of exploring ethical considerations during the design process.[49] Infrastructuring may be a viable first step toward crafting technology use cases that explore social bias to mitigate unethical AI development.

**Organizational**

To allow for design after design, a specific organizational structure that supports adaptability needs to be in place. The discipline has made strides, particularly by using design thinking, in creating economic benefits for businesses. In the context of AI, we contend that a viable first step would be adopting the principles of organizational ambidexterity.[50] An increasingly popular construct for achieving long-term firm survival,[51] organizational ambidexterity is defined as "the ability to simultaneously pursue both incremental and discontinuous innovation. … Hosting multiple contradictory structures, processes, and cultures within the same firm."[52] Organizational ambidexterity allows companies to simultaneously manage current business demands and adapt to environmental changes.[53] We believe design-led ambidexterity is particularly well suited in a context where a solution is never fully complete but in a state of continuous reconfiguration based on new insights.

---

46 Joyce Chou, Oscar Murillo, and Roger Ibars, "What the Kids' Game 'Telephone' Taught Microsoft about Biased AI," *Co.Design* (October 12, 2017).

47 Jeanne Liedtka, "Perspective: Linking Design Thinking with Innovation Outcomes through Cognitive Bias Reduction," *Journal of Product Innovation Management* 32, no. 6 (2015): 925–38.

48 Chou, Murillo, and Ibars, "What the Kids' Game 'Telephone' Taught Microsoft."

49 Peter Lloyd, "Ethical Imagination and Design," *Design Studies* 20, no. 2 (2018): 154–68.

50 Charles A. O'Reilly III and Michael L. Tushman, "Organizational Ambidexterity: Past, Present, and Future," *Academy of Management Perspectives* 27, no. 4 (2013): 324–38.

51 Jana Oehmichen, Mariano L. M. Heyden, Dimitrios Georgakakis, and Henk W. Volberda, "Boards of Directors and Organizational Ambidexterity in Knowledge-Intensive Firms," *International Journal of Human Resource Management* 28, no. 2 (2017): 283–306.

Ipso facto organizational adaptability has implications for the funding model of design firms. The billable-hour funding model is standard practice for design agencies and consultancies. However, a model is needed that could address the new market situation in which AI-powered solutions continuously evolve and each of their incarnations has to be monitored and adapted when needed to ensure the delivered solutions are aligned with human interests. We contend that a retainer funding model could prove to be a better fit for design-client management in such situations as it allows agencies to bill a client each annum (or quarterly) to "retain" their services. The design agency can access this retainer funding to complete scheduled audits of AI-powered products and services on a continuous basis. Where the billable hour model is required again, perhaps for major updates or specific projects, it can be reintroduced as necessary.

**Technological**

To address this part of the infrastructure, designers need to understand the technology (particularly its core—the algorithm) they are going to work with and be able to "converse" with it. Much like one needs to know programming technologies such as HTML and CSS to translate their design into a digital product (a website), designers need to understand the "new material" they will use. Progress is already being made in that direction with multiple emerging Meetup groups between AI developers and designers around the world, companies and consultancies publishing their principles on how to design with AI (e.g., Google, IDEO, Microsoft, Fjord), and speculative design exhibitions to give shape to AI's possible incarnations (e.g., IDEO's exhibition HyperHuman).

However, AI algorithms are often a mystery, even to their own developers. This posits a question: how do we understand something that is inherently opaque? We believe initial answers might be hidden in design cognition. At first glance, algorithmic logic appears to be most detached from the discipline of design. Practically, this may be the case. Theoretically, formulating algorithms and the nature of design cognition are closer than they seem. One of the fundamental paradigms of design methodology and consequently cognition—design as a rational problem solving process introduced by Newell and Simon in the early 1970s—originated within the field of AI.[54] In a turn of events, the disciplines meet again with AI-powered solutions entering wicked problem domains. We believe that principles developed in design cognition could be translated back to devising algorithms. However, this proposition requires the ability to overcome the problems with the widely criticized areas of Newell and Simon's paradigm, such as failing to account for the action-oriented, often implicit knowledge associated with design.[55] Further research is needed to reach a deeper understanding of design cognition and its possible future incarnations and implications for AI.

**Conclusion**

This article sheds light on the deeper consequences of AI development for the discipline of design. We have argued that to ensure beneficial outcomes of AI, design and its underlying rationale can create a new paradigm for the development of AI-powered solutions. We identify that there are fragments of methodological readiness for AI within the discipline: human-centeredness, co-performance, and an attention to ethics. We have brought these methodological developments together around the notion of infrastructure to propose how to design with the new subject matter of AI. This is especially important if ML will be the most important way to improve not only user experience but also the way such solutions impact humans and communities.

The strength and novelty of our proposition stems from the proposal to design infrastructures by understanding the interrelations and implications of its social, organizational, and technological elements. Further research is needed on each element and their potential symbiosis. We believe one way to do so is by continuously setting up small design experiments that can prototype infrastructures. The designer begins using current knowledge of prototyping (experiences, product service systems, organizations) to continuously generate and test assumptions. The designer becomes responsible for shaping solutions beyond the primary function, anticipating and evaluating new horizontal functions in all three elements of infrastructure: society, organizations, and technology. In development, customers and stakeholders are engaged to define the problem and create shared value. In production, manufacturers and material scientists are connected with programmers and system architects. Furthermore, the discipline should envision possible types of information that can be collected and how processing and sharing that information might enrich the user's life beyond a current set of lived and potential situations.

---

54 Allen Newell and Herbert Alexander Simon, *Human Problem Solving* (Englewood Cliffs, NJ: Prentice Hall, 1972); Kees Dorst, "The Problem of Design Problems," *Expertise in Design* (2003): 135–47.

55 Dorst, "The Problem of Design Problems."

The notion of co-performance allows for new arrangements between users and objects that are powered by AI. What is promising is the collaborative efforts of industry and academia to engage in and share scientific breakthroughs regarding AI. Our proposition of harnessing infrastructure to approach AI is a step forward. Many more steps will be required to reach methodological clarity. Clearly, the age of prevalent adoption of AI-powered solutions is here, and design can play a vital role in creating solutions that grow with people, not against them.